\documentclass[aps,prl,twocolumn,amsmath,amssymb,showpacs]{revtex4}
\usepackage{epsfig,float}
\usepackage{graphicx}
\usepackage{dcolumn}
\usepackage{morefloats}
\usepackage{color}
\usepackage{slashed}
\usepackage{bbm}

\begin{document}

\title{Decoding the riddle of $Y(4260)$ and $Z_c(3900)$}

\author{Qian Wang$^1$\footnote{{\it Email address:} q.wang@fz-juelich.de},
Christoph Hanhart$^1$\footnote{{\it Email address:}
        c.hanhart@fz-juelich.de}, Qiang Zhao$^2$\footnote{{\it Email address:}
        zhaoq@ihep.ac.cn} }

\affiliation{
       $^1$ Institut f\"{u}r Kernphysik and  Institute for Advanced Simulation,
          Forschungszentrum J\"{u}lich, D--52425 J\"{u}lich, Germany\\
       $^2$ Institute of High Energy Physics and Theoretical Physics Center for Science Facilities,
        Chinese Academy of Sciences, Beijing 100049, China}

\begin{abstract}

The observation of $Z_c(3900)$ by the BESIII collaboration in the
invariant mass spectrum of $J/\psi\pi^\pm$ in $e^+e^-\to
J/\psi\pi^+\pi^-$ at the center of mass 4.260 GeV suggests the
existence of a charged $\bar D D^*+D\bar D^*$  molecular state with
$I(J^P)=1(1^+)$, which could be an isovector brother of the famous
$X(3872)$ and an analogue of $Z_b(10610)$ claimed by the Belle
Collaboration. We demonstrate that this observation provides strong
evidence that the mysterious $Y(4260)$ is a $\bar{D}D_1(2420)
+D\bar{D}_1(2420)$ molecular state. Especially, we show that the
decay of this molecule naturally populates low momentum  $\bar DD^*$
pairs and leads  unavoidably to a cusp at the $\bar DD^*$ threshold.
We discuss the signatures that distinguish such a $\bar DD^*$ cusp
from the presence of a true resonance.

\end{abstract}

\date{\today}






\maketitle

During the past years, the experimental observation of a large
number of so-called $X$, $Y$, $Z$ states has initiated tremendous
efforts to unravel their nature beyond the conventional quark model.
Especially, the confirmation of signals in charged channels would be
a direct evidence  for exotic states. For instance, the Belle
Collaboration reported signals for $Z(4430)$ in $\psi'\pi^\pm$, and
$Z_1(4050)$ and $Z_2(4250)$ in $\chi_{c1}\pi^{\pm}$ in $B$ meson
decays~\cite{belleZ+}. However, an enhancement in the same mass
range was interpreted as a reflection by BaBar~\cite{babarZ+}. The
more recent experimental results for charged bottomonium states
$Z_b(10610)$ and $Z_b(10650)$, located close to the $\bar BB^*$ and
$\bar B^*B^*$ thresholds, respectively, by the Belle Collaboration~\cite{Zb}
seem to be the first strong evidence for QCD ``exotics" in the heavy
quark sector. In this context the recent report of an enhancement in
the $J/\psi \pi^\pm$ invariant mass distribution around 3900 MeV,
right at the $\bar DD^*$ threshold, by the BESIII
collaboration~\cite{BESIIIZ+} clearly reinforces the existence of
such an unusual phenomenon. This state, called $Z_c(3900)$ below,
might be an isovector partner of the well established $1^{++}$
isoscalar $X(3872)$~\cite{Beringer:1900zz}, but with
$I^G(J^{PC})=1^+(1^{+-})$ for the neutral state.

In this work we demonstrate that, if $Y(4260)$ is a
$\bar{D}D_1+c.c.$ molecule (below we use $\bar{D}D_1$ as a short
notation), the appearance of an enhancement around 3900 MeV in the
$J/\psi \pi$ invariant mass distribution can be shown to be natural.
Here $D_1$ refers to the narrow axial vector $D_1(2420)$
($\Gamma=27\pm 3$ MeV) with $I(J^P)=\frac
12(1^+)$~\cite{Beringer:1900zz}. In this sense the observation of
the charged $Z_c(3900)$ state by BESIII in $Y(4260)\to J/\psi\pi\pi$
provides a very strong evidence for the molecular nature of
$Y(4260)$. We also discuss whether the observed enhancement can be
interpreted purely as a cusp or whether the inclusion of explicit
poles in the $\bar DD^*$ system is necessary. Before we go into
details of our calculations we first briefly review the status of
$Y(4260)$.

The most mysterious
fact about $Y(4260)$ is not that its mass does not agree to what is
predicted by the potential quark model. Instead, as a charmonium
state with $J^{PC}=1^{--}$, it is only ``seen" as a bump in its two
pion transitions to $J/\psi$, but not in any open charm decay
channel like $\bar{D}D$, $\bar{D}D^*+c.c.$, $\bar{D}^*D^*$ and
$\bar{D}_sD_s^*$, or other tens of measured channels. In fact, the
cross section lineshapes of the $e^+ e^-$ annihilations into
$D^{(*)}$ meson pairs appear to have a dip at its peak mass 4.26 GeV
instead of a bump.

In the vector sector one should recognize that
the $\bar{D}D_1(2420)$ is the first open charm relative $S$-wave
channel coupled to $J^{PC}=1^{--}$, and the nominal
threshold is only about 29 MeV above the location of the $Y(4260)$.
The possibility that $Y(4260)$ may be a bound system of $\bar{D}D_1$
and $\bar D^*D_0$ was investigated in
Refs.~\cite{swanson,Close:2010wq,Ding:2008gr,Li:2013bca}. It should
be stressed, however, that broad components (the widths of $D_0$ and
$D_1(2430)$ are as large as 300 MeV) can not produce narrow
resonances~\cite{ourcomment}. We thus do not consider the $\bar
D^*D_0$ or $\bar D^*D_1(2430)$ component here, but focus on the
assumption that the $Y(4260)$ is a bound system of only
$\bar{D}D_1(2420)$.

In the literature many other solutions
were proposed for the $Y(4260)$ (see Ref.~\cite{Brambilla:2010cs}
for a recent review). Based on the data for the $\pi\pi$ spectrum of
$Y(4260)\to J/\psi\pi\pi$~\cite{Lees:2012cn}, Dai {\it et al.}~\cite{Dai:2012pb}
concluded that the even lower threshold $\chi_{c0}\omega$ would have
the largest coupling to $Y(4260)$, while the $\bar{D}D_1$ coupling
to $Y(4260)$ turned out to be negligible. However, the observation
of the enhancement at the $\bar DD^*$ threshold in the $J/\psi\pi$
invariant mass spectrum actually rules out such a scenario and
suggests that the underlying dynamics should be more sensitive to
the $\bar{D}D_1$ threshold.

In this work we do not try to identify the mechanisms that lead to
the formation of the $Y(4260)$ as a molecular state, but study the
consequences of the assumption that  this state is dominantly a $\bar{D}D_1$
bound system. We argue that the interpretation of $Y(4260)$ as a
relative $S$-wave $\bar{D}D_1$ system is able to accommodate nearly
all the present observations for $Y(4260)$. Especially, its absence
in various open charm decay channels mentioned above and the
observation of $Z_c(3900)$ in $Y(4260)\to J/\psi\pi\pi$ can be
naturally understood.

The heavy quark spin symmetry also implies the presence of a
$\bar{D}^* D_1$ and $\bar{D}^*D_2$ component in the wave function.
However, they will be neglected here since their corresponding
thresholds are almost 200 MeV above the mass of $Y(4260)$.
Meanwhile, we note that in a more microscopic treatment there should
be heavier spin partners of $Y(4260)$ that should contain the
mentioned constituents prominently. We will briefly come back to
this issue at the end of this paper.

\begin{figure}[t] \vspace{0.cm}
\begin{center}
\hspace{2cm}
\includegraphics[scale=0.6]{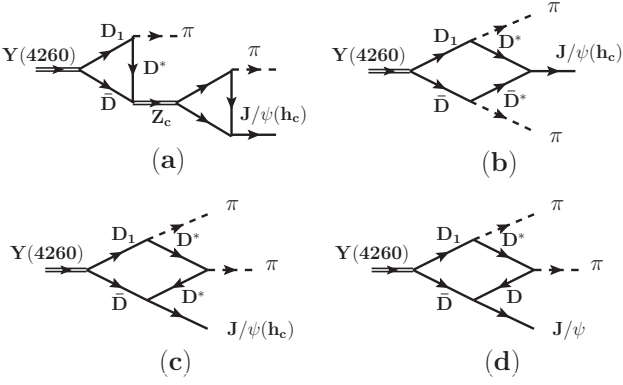}
\caption{The Feynman diagrams for $Y(4260)\to \bar{D}D_1+c.c.\to
J/\psi \pi\pi$ and $h_c\pi\pi$ considered in this work.}
 \label{fig-1}
\end{center}
\end{figure}

Based on the above picture, the pertinent diagrams to be calculated are shown in
Fig.~\ref{fig-1}. The central goal of our study is to pin down the
structure of $Y(4260)$ and identify the quantitative importance of
the $\bar DD^*$ cusp from the loop diagrams (b)-(d) in order to
understand whether the $\mbox{BESIII}$ spectra call for the
additional inclusion of an explicit pole diagram as depicted in
diagram (a).

It should be stressed that the vicinity of the $\bar{D}D_1$
threshold to 4.26 GeV favors the formation of low momentum $\bar
DD^*$ pairs, which may lead to the formation of the $Z_c(3900)$ bound
systems. The reason is that the $\bar{D}D_1$ intermediate state as
well as the $\bar DD^*$ intermediate state can be simultaneously
close to their mass shells --- all with relative $S$ waves. This
gives rise to a triangle singularity studied in a different context
in Refs.~\cite{Wu:2011yx,Wu:2012pg}. Such a two-cut condition
strongly enhances the corresponding matrix elements. In this sense
the $\bar{D}D_1$ intermediate system provides an ideal doorway state
for a low momentum $\bar DD^*$ system. There are a series of
$S$-wave open charm thresholds with $J^{PC}=1^{--}$ around 4.26 GeV,
i.e. $\bar{D}D_1(2420)$, $\bar{D}D_1(2430)$, $\bar D^* D_0$,
$\bar{D}_{s0}D_s^*$, $\bar{D}_s D_{s1}$, and $\bar{D}^*D_1(2420)$.
However, all of them, except for $\bar{D}D_1(2420)$, are either far
away from the observed physical $Y(4260)$, or too broad to make a
bound state~\cite{Beringer:1900zz}. Thus, we do not take them into
account explicitly here.

Note that in order to formulate the problem as an effective field
theory the power counting of Refs.~\cite{Guo:2010ak,Cleven:2013sq}
needs to be adapted to the present situation. We leave this to be
reported in a subsequent work and focus here on a more
phenomenological investigation.

A complete calculation also needs the inclusion of the $\pi\pi$
final state interaction (FSI) for which we adopt a parametrization scheme with coupled channel unitarity~\cite{Dai:2012pb,Bugg:1996ki}.
Since the pion pairs are in the isoscalar channel and the invariant mass of the pion pairs covers a range from the two pion threshold to more than 1 GeV,  the $S$-wave $\pi\pi$ FSI is expected to play an important role in this region. We can isolate the $\pi\pi$ S-wave contributions in the $\pi\pi$ center of mass frame as follows:
\begin{eqnarray}
\mathcal{M}=\mathcal{M}_S+\mathcal{M}_{non-S}.
\end{eqnarray}
After including the $\pi\pi$ S-wave FSI, the amplitude becomes
\begin{eqnarray}
\mathcal{M}^\prime\equiv \mathcal{M}_S\alpha(s_{\pi\pi}) \mathcal{T}_{\pi\pi\to\pi\pi}+\mathcal{M}_{non-S} \ ,
\end{eqnarray}
where $\mathcal{T}_{\pi\pi\to\pi\pi}$ is the $\pi\pi$ elastic scattering amplitude \cite{Bugg:1996ki} with the $K\bar{K}$ threshold appropriately considered and $\alpha(s_{\pi\pi})$ is a polynomial function of the $\pi\pi$ invariant mass squared $s_{\pi\pi}$ \cite{Dai:2012pb}
\begin{eqnarray}
\alpha(s_{\pi\pi})=\frac{c_1}{s_{\pi\pi}-m_\pi^2/2}+c_2+c_3 s_{\pi\pi} \ ,
\end{eqnarray}
where the Adler zero pole $m_\pi^2/2$ is present in order to cancel the Adler zero hidden in $\mathcal{T}_{\pi\pi\to\pi\pi}$. The following parameters, $c_1=0.23$ GeV$^2$, $c_2=-1.07$ and $c_3=1.15$ GeV$^{-2}$ are adopted for a reasonable description of the experimental data.

Since $Z_c(3900)$ has the same quantum numbers as $Z_b$, most interactions
needed for this work can be taken from Ref.~\cite{Cleven:2013sq}.
Here we only present the interactions between a $P$-wave charmed
meson and other fields since they play a crucial role in this work.
The heavy quark spin symmetry allows heavy mesons to form spin
doublet super fields distinguished by their light degrees of freedom
$s_l=s_q+l$ with $s_q$ the light quark spin and $l$ the orbital
angular momentum. For $l=1$ it can be classified into two super
fields~\cite{Casalbuoni:1996pg,Colangelo:2005gb}. The interaction
terms relevant for this work read
\begin{small}
\begin{eqnarray}
\mathcal{L}_Y&=&iy(\bar{D}_a^\dag Y^i
D_{1a}^{i\dag}-\bar{D}_{1a}^{i\dag} Y^i D_a^\dag)+h.c.,\\\nonumber
\mathcal{L}_{D_1}&=&i\frac{h^\prime}{f_\pi}
[3D_{1a}^i(\partial^i\partial^j\phi_{ab})D^{*\dag
j}_b-D_{1a}^i(\partial^{j}\partial^j\phi_{ab})D_b^{*\dag
i}\\& +&3\bar{D}_a^{*\dag
i}(\partial^i\partial^j\phi_{ab})\bar{D}_{1b}^j-\bar{D}_a^{*\dag
i}(\partial^j\partial^j\phi_{ab})\bar{D}_{1b}^i]+h.c.
\end{eqnarray}
\end{small}
where $D$ ($D^\dag$) and $\bar{D}$ ($\bar{D}^\dag$) contain the
annihilation (creation) operators for  $c\bar{q}$ and $\bar{c}q$
fields, respectively.  The analogous conventions are applied to
$D^*$ and $D_1$. The interactions contain five coupling constants in
total. Since we focus on the shape of the invariant mass
distributions only, all couplings that are common to all diagrams
are not relevant. In this sense the only free parameters that
influence the invariant mass distributions are the mass of $Z_c(3900)$ and
the coupling constant $g_Z$ for $Z_c\bar{D}D^*$. Meanwhile, we adopt a Breit-Wigner propagator for the  $Z_c(3900)$.

For simplicity in this exploratory study we treat $Y(4260)$,
$D_1(2420)$, $D$ and $D^*$ as stable states. Their masses are taken
from the Particle Data Group~\cite{Beringer:1900zz}.
By assuming that $Y(4260)$ is dominated by the $\bar{D}D_1(2420)$
molecule component, we can estimate the $YDD_1$ coupling by
Weinberg's compositeness theorem~\cite{Weinberg:1965zz,
Baru:2003qq}, i.e. $y^2/4\pi\equiv 4(m_D+ m_{D_1})^{5/2}
\sqrt{2\delta E/m_D m_{D_1}}\simeq 17$ GeV$^2$. This predicts the
dominant decay of $Y(4260)\to \bar{D}D^*\pi+c.c.$ via the
intermediate state $\bar{D}D_1(2420)+c.c.$ of which the partial
width is larger than 40 MeV. This value is consistent with the total
width measured for $Y(4260)$.
At this moment, we do not pursue a perfect fit of the experimental data but demonstrate the importance of the proposed mechanisms in the description of the qualitative feature of the data. In fact, with only one parameter, $g_Z=1 ~\mathrm{GeV}^{-1/2}$ which gives the $Z_c\to DD^*$ branching ratio of about $20\%$,
the data can be described well in our scenario.

\begin{figure} \vspace{0.cm}
\vspace{-1cm}
\includegraphics[scale=0.17]{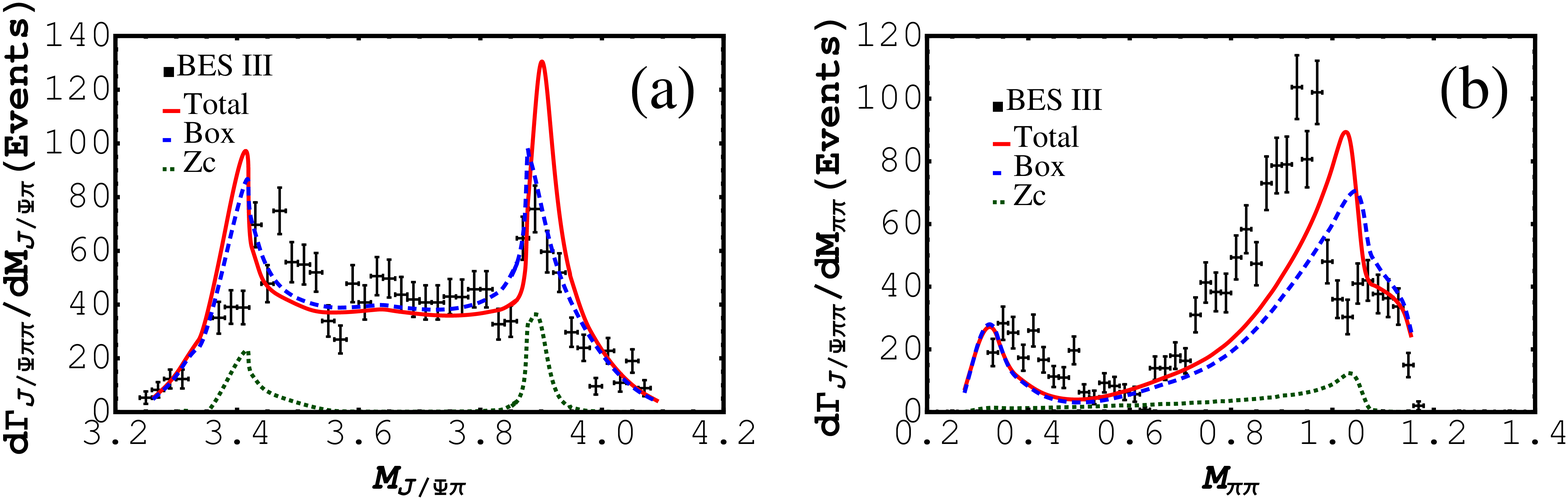}
\vspace{-5cm}
\caption{The invariant mass spectra for (a) $J/\psi\pi$ and (b)
$\pi\pi$ in $Y(4260)\to J/\psi\pi\pi$. The solid, dashed, and
dotted lines stand for the results of the full
calculation, box diagrams, and triangle diagram with the $Z_c(3900)$ pole,
respectively.}
 \label{fig-2}
\end{figure}

The numerical results for the $J/\psi\pi$ and $\pi\pi$
invariant mass spectra from $Y(4260)\to J/\psi\pi\pi$ are shown in
Fig.~\ref{fig-2}.
The dashed lines denote the results from the box diagrams, i.e. Fig.~\ref{fig-1} (b)-(d), while the dotted lines denote the exclusive contributions from the $Z_c(3900)$ pole diagrams, i.e. Fig.~\ref{fig-1} (a). In our case the box diagrams play a role as background terms in respect to the $Z_c(3900)$ pole diagrams. However, as shown by the dashed lines, an explicit enhancement around 3.9 GeV in the $J/\psi\pi$ spectrum can be produced because of the nearly on-shell two-cut condition.
The solid lines in Fig.~\ref{fig-2} show the result of the sum of
all diagrams in Fig.~\ref{fig-1}. The $Z_c(3900)DD^*$ coupling is chosen
in order to reproduce the qualitative features of the data. The inclusion of the $Z_c(3900)$ actually broadens and enhances the
$\bar{D}D^*$ threshold enhancement and gives rise to the detailed structures in the $\pi\pi$ invariant spectrum. There are signatures that can be identified for the $\pi\pi$ production mechanism. An analysis of the relative partial waves between the two pions suggests that in addition to the $S$ wave other higher partial waves, such as $D$ wave, are also contributing to the $\pi\pi$ productions. Meanwhile, the dominance of the relative $S$ wave $\pi\pi$ in the lower invariant mass region results in the broad bump above the $\pi\pi$ threshold and flattened dip around 0.5-0.6 GeV. This structure is driven by the box diagrams after the $\pi\pi$ FSI is properly included. As a contrast, the exclusive contributions from the $Z_c(3900)$ pole amplitude do not produce an obvious structure in the $\pi\pi$ spectrum. It is essential to recognize that the dip structure around 1 GeV in Fig.~\ref{fig-2} (b) is due to the presence of the $K\bar{K}$ threshold in the $\pi\pi$ FSI. The exact $K\bar{K}$ threshold should be located around 0.986 GeV. However, the data show that the dip position is slightly shifted to be higher than 1 GeV. In our calculation such a shift is due to the contributions from higher partial waves.

For the $J/\psi\pi$ spectrum, one can see that even without the
explicit inclusion of $Z_c(3900)$,
 two structures appear  at the same masses in the $J/\psi
\pi$ spectrum as in the BESIII data.  Note that by charge
conjugation invariance the $J/\psi \pi^-$ spectrum is identical with
that for $J/\psi \pi^+$. The cusp at $M_D+M_{D^*}\simeq 3.876$ GeV
in Fig.~\ref{fig-2}(a) marks the $\bar DD^*$ threshold. The lower
bump between 3.4 and 3.6 GeV in the $J/\psi\pi^+$ invariant mass
distribution comes from the interference between the conjugate
diagrams where either a $\pi^-$ or a $\pi^+$ is emitted first.
Therefore, the lower bump is simply a reflection of the narrow
structure at 3.9 GeV.
One also notices that the box diagrams are the main contributions as a background to the $J/\psi\pi$ spectrum away from the $Z_c(3900)$ pole.

\begin{figure} \vspace{0.cm}
\vspace{-1cm}
\includegraphics[scale=0.17]{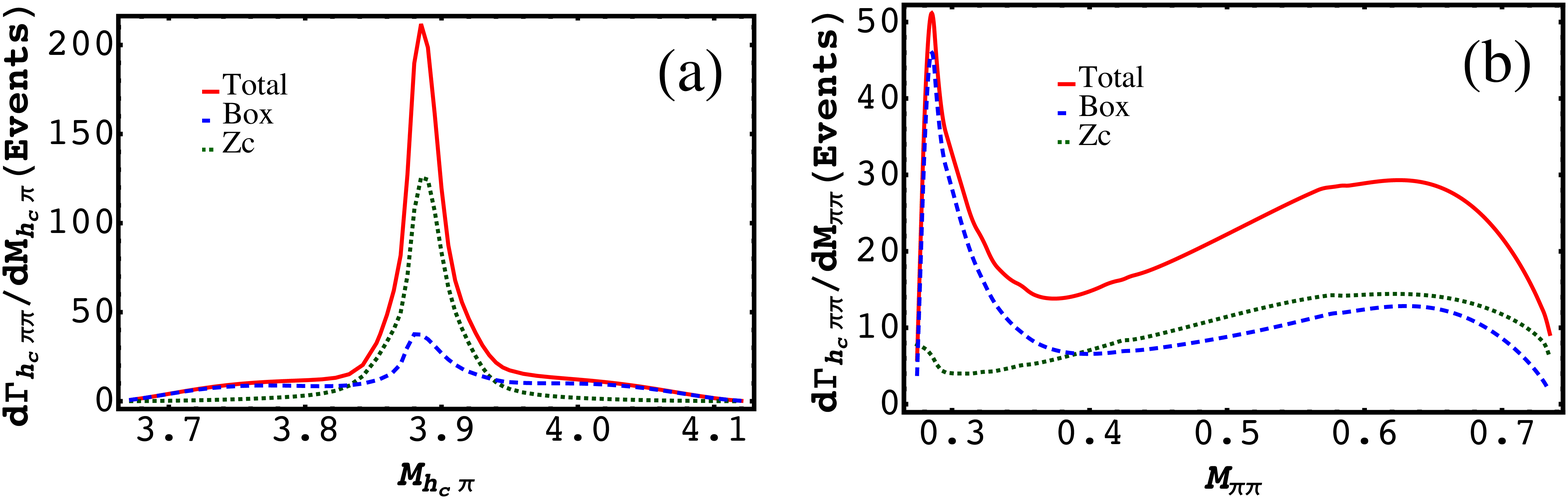}
\vspace{-5cm}
\caption{The invariant mass spectra for (a) $h_c\pi$ and (b)
$\pi\pi$ in $Y(4260)\to h_c\pi\pi$. The notations are the same as in
Fig.~\ref{fig-2}. }
 \label{fig-3}
\end{figure}

In Fig.~\ref{fig-1}, we also include the production channel for
$Y(4260)\to h_c\pi\pi$. Similar to the discussions of $Z_b\to
h_b\pi$, this channel is ideal for disentangling the molecular
nature of the intermediate $Z_c(3900)$: The power counting analysis in
Ref.~\cite{Cleven:2013sq} shows that the triangle transition $Z_b\to
h_b(mP)\pi$ is not suppressed compared to $Z_b\to \Upsilon(nS)\pi$,
although the decay is via a $P$-wave. This explains why the
branching ratios for $Z_b\to h_b(mP)\pi$ are compatible with those
for $Z_b\to \Upsilon(nS)\pi$. A similar phenomenon occurs here. In addition,
 higher loop contributions are
suppressed in $Z_b\to h_b(mP)\pi$, while they are not suppressed for
$Z_b\to \Upsilon \pi$~\cite{Cleven:2013sq}. The analogous pattern is
expected for the $Z_c$ decays.

In Fig.~\ref{fig-3}, we present our prediction for the invariant
mass spectra of $h_c\pi$ and $\pi\pi$ in $Y(4260)\to h_c\pi\pi$ including the $\pi\pi$ FSI.
Similar
to the $J/\psi\pi\pi$ channel, a very pronounced peak right at the
$\bar DD^*$ threshold appears in the $h_c\pi$ invariant spectrum.
Interestingly, due to the limited phase space in this decay channel,
its kinematic reflection is significantly shifted.  It is located at
higher invariant masses and even submerged by the $\bar{D}D^*$ threshold enhancement.
With the parameters fixed in $Y(4260)\to J/\psi\pi\pi$, it shows that the enhancement at 3.9 GeV produced by the box diagrams is not as significant as that by the explicit $Z_c(3900)$ pole. Meanwhile, the $\pi\pi$ spectrum at lower mass regions is sensitive to the underlying dynamics. This feature is quite different from that observed in the $J/\psi\pi\pi$ channel.
Experimental data for this channel will allow a clear evidence for the request or elimination of
the $Z_c(3900)$ resonance contribution. Also, different from the $J/\psi\pi\pi$ transition, the predicted $\pi\pi$ spectrum shown in Fig.~\ref{fig-3} (b) does not have the $\bar{K}K$ threshold discontinuation due to the limited phase space. We find that the $S$-wave $\pi\pi$ amplitude still plays an important role in the $h_c\pi\pi$ channel with the $\pi\pi$ in a $P$ wave relative to the recoiled $h_c$.

We stress that in order to understand the gross features of the data
the diagrams in Fig.~\ref{fig-1} can also be replaced completely or
in parts by tree level diagrams (see, e.g.,
Refs.~\cite{Chen:2012yr,Chen:2011pv}). Then, however, the underlying
dynamics with the interplay of the $\bar DD^*$ and $\bar DD_1$ cuts
will be lost. We reiterate that it is the $\bar{D}D_1$ molecule
nature of $Y(4260)$ that provides a natural explanation for the
appearance of $Z_c(3900)$ in the $Y(4260)$ decays, and for other
detailed features of the spectra.

In this work we propose that $Y(4260)$ is dominantly a
$\bar{D}D_1(2420)$ molecule and identify a unique mechanism, namely
the presence of two-cut condition, which plays an essential role in
$Y(4260)\to J/\psi\pi\pi$ and $h_c\pi\pi$. We demonstrate that
without introducing any drastic assumption, the molecular nature of
$Y(4260)$ as a bound state of $\bar{D}D_1$ can naturally explain the
observation of an enhancement around $3.876$ GeV in the $J/\psi\pi$
invariant mass spectrum. The reflection of this peak matches the experimental data well and provides  strong
evidence for the molecular nature of $Y(4260)$. We also demonstrate
that for a more detailed description of the data the need for an explicit $Z_c(3900)$ pole seems to be necessary.

We stress the following important consequences of this prescription
that will be reported in a subsequent paper:

i) Although the nominal $\bar{D}D_1(2420)$ threshold is higher than the mass
of $Y(4260)$, the threshold of $\bar{D}D^*\pi$ (then decaying to
$\bar{D}D\pi\pi$) is much lower. Dominant decays of $Y(4260)$ into
$\bar{D}D^*\pi$ should be regarded as a natural consequence and
would explain the large deficit in width between the total width and
its decays into $J/\psi\pi\pi$ and $J/\psi K\bar{K}$. We also
predict an asymmetric spectral shape for the decay of $Y(4260)$ into
$\bar{D}D^*\pi$. This can be clarified by an energy scan around the
nominal $Y$ mass (analogous to what happens to the shape of
$Y(4660)$ when being viewed as $\psi'f_0(980)$
molecule~\cite{our4660}).

ii) A possible reason why $Z_c(3900)$  does not show up as significantly as the $X(3872)$ in $B$ decays is that the $X(3872)$ might
be produced via its small $\bar cc$ component which is absent in
$Z_c$. Here, $Z_c(3900)$ does appear as a result of the proposed
molecular nature of $Y(4260)$, which actually allows the $S$-wave
$\bar DD^*$ pairs to be copiously produced.

iii) One might also expect a cusp or possible resonance structure at
the $\bar{D}^*D^*$ threshold in the decay of $\psi(4415)$, if this
state is assumed to be a molecule driven by the nearby
$\bar{D}^*D_1$ threshold.

iv) The exact mechanism at work here might also be the reason for
the appearance of the $Z_b$ and $Z_b'$ states in $\Upsilon(5S)$
decays, if we assume that $\Upsilon(5S)$ has a sizeable $\bar{B}B_1$
component. Note that the relative $S$-wave threshold for
$\bar{B}B_1$ is only about 120 MeV above the $\Upsilon(5S)$ mass.

The authors thank F.-K. Guo, G. Li, U.-G. Meissner, C.-Z. Yuan, and
B.-S. Zou for useful discussions. A special acknowledgement is to Thomas Hahn for his help on the use of the LoopTool package. This work is supported, in part,
by the National Natural Science Foundation of China (Grant Nos.
11035006 and 11121092), the Chinese Academy of Sciences
(KJCX3-SYW-N2), the Ministry of Science and Technology of China
(2009CB825200), and DFG and NSFC funds to the Sino-German CRC 110.

\end{document}